\documentstyle[aas2pp4]{article}

\lefthead{OROSZ ET AL.}
\righthead{PG1224+309}

\begin{document}

\title{The Post-Common Envelope and Pre-Cataclysmic Binary
PG 1224+309\protect\footnote{Based in part on observations obtained at
the Michigan-Dartmouth-MIT Observatory.}  }
\author{Jerome A. Orosz\altaffilmark{2}, Richard A. Wade, and 
Jason  J. B. Harlow\altaffilmark{2}}
\affil{Department of Astronomy \& Astrophysics, The
Pennsylvania State University, 525 Davey Laboratory, University Park, PA
16802-6305\\
orosz@astro.psu.edu, wade@astro.psu.edu, harlow@astro.psu.edu}

\author{John R. Thorstensen and Cynthia J. Taylor}
\affil{Department of Physics and Astronomy,
6127 Wilder Laboratory, Dartmouth College,
Hanover, NH 03755-3528\\
thorstensen@dartmouth.edu, Cynthia.J.Taylor@dartmouth.edu}

\and

\author{Michael Eracleous\altaffilmark{3,4}}
\affil{Department of Astronomy, University of California, Berkeley, CA 94720}

\altaffiltext{2}{Visiting Astronomer at Kitt Peak National 
Observatory (KPNO), which is operated by AURA, Inc., under a cooperative 
agreement 
with the National Science Foundation.}
\altaffiltext{3}{Hubble Fellow.}
\altaffiltext{4}{Current address: 
Department of Astronomy and Astrophysics, The 
Pennsylvania State University, 525 Davey Lab, University Park, PA 16802,
electronic mail: mce@astro.psu.edu}

\begin{abstract}
We have made extensive spectroscopic and photometric observations of
PG 1224+309, a close binary containing a DA white dwarf primary and an
M4+ secondary.  The H$\alpha$ line is in emission due to irradiation
of the M-star by the hot white dwarf and is seen to vary around the
orbit. From the radial velocities of the H$\alpha$ line we derive a
period of $P=0.258689\pm 0.000004$ days and a semi-amplitude of
$K_{{\rm H}\alpha}=160\pm 8$ km s$^{-1}$.  We estimate a correction
$\Delta K=21\pm 2$ km s$^{-1}$, where $K_{\rm M}=K_{{\rm
H}\alpha}+\Delta K$.  Radial velocity variations of the white dwarf
reveal a semi-amplitude of $K_{\rm WD}=112\pm 14$ km s$^{-1}$.  The
blue spectrum of the white dwarf is well fit by a synthetic spectrum
having $T_{\rm eff}=29,300$ K and $\log g=7.38$.  The white dwarf
contributes 97\% of the light at 4500~\AA\ and virtually all of the
light blueward of 3800~\AA.  No eclipses are observed. The mass
inferred for the white dwarf depends on the assumed mass of the thin
residual hydrogen envelope: $0.40\le M_{\rm WD}\le 0.45\,M_{\sun}$ for
hydrogen envelope masses of $0\le M_H \le 4\times 10^{-4}\,M_{\odot}$.
We argue that the mass of the white dwarf is closer to
$0.45\,M_{\odot}$, hence it appears that the white dwarf has a
relatively large residual hydrogen envelope.  The mass of the M-star
is then $M_{\rm M}=0.28\pm 0.05\,M_{\odot}$, and the inclination is
$i=77\pm 7^{\circ}$. We discuss briefly how PG 1224+309 may be used to
constrain theories of close binary star evolution, and the past and
future histories of PG 1224+309 itself.  The star is both a
``post-common envelope'' star and a ``pre-cataclysmic binary'' star.
Mass transfer by Roche-lobe overflow should commence in about
$10^{10}$ yr.
\end{abstract}

\section{Introduction}

PG 1224+309 was cataloged as a UV-excess object in the Palomar-Green
(PG) survey (Green, Schmidt, \& Liebert 1986\markcite{green}).  It
also appears in the Tonantzintla survey as Ton 617 (Iriarte \& Chavira
1957\markcite{Ira}; Chavira 1958\markcite{Chav}) and the Case
low-dispersion northern sky survey as CBS 60 (Sanduleak \& Pesch
1984\markcite{sand}).  Its mean magnitude and colors are $V=16.164$,
$B-V=-0.065$, $V-R=0.038$, and $R-I=0.312$ (this work).  This object
was studied by Ferguson, Green, \& Liebert (1984)\markcite{FGL} who
were searching for cataclysmic variable-like stars in the PG survey.
Ferguson et al.\ (1984)\markcite{FGL} showed that PG 1224+309 was a
binary, identified the hot star as a DA white dwarf with $T_{\rm
eff}=28,000$~K and $\log g=7.7\pm 0.8$, and classified the cool star
as an M2 dwarf based on its colors.  Refined values of $T_{\rm
eff}=29,300\pm 1000$~K and $\log g=7.38\pm 0.1$ for the white dwarf
were later provided by J.\ W.\ Liebert \& P.\ Bergeron
(1997)\markcite{lieb}.  PG1224+309 was not discussed again in the
literature for thirteen years.  Orosz, Wade, \& Harlow
(1997\markcite{orosz1}) looked for radial velocity variations in a
number of stars that were characterized as composite spectrum binaries
by Ferguson et al.\ (1984)\markcite{FGL}, and found for PG 1224+309
that the radial velocity of the M star changed by $83\pm 11$ km
s$^{-1}$ between two observations $\approx 0.9$ days apart.  Orosz et
al.\ (1997\markcite{orosz1}) also showed that PG 1224+309 has a strong
and variable H$\alpha$ emission line.  Photometric observations by
S. Bell (private communication 1996) showed that the light varies
smoothly on a timescale of hours.

\begin{deluxetable}{ccccc}
\tablewidth{0pt}
\tablecolumns{5}
\tablecaption{Log of Spectroscopic Observations}
\tablehead{
\colhead{Start\tablenotemark{a}} & \colhead{Finish\tablenotemark{a}} & 
\colhead{$N$} 
& \colhead{Setup} & \colhead{Detector} \\
}
\startdata
447.03 & 453.05 & 8  &MDM 2.4+mod & Tek1024 \nl
486.89 & 491.96 & 18 &MDM 1.3+MkIII & Tek1024 \nl
494.94 & 498.90 & 8  &MDM 2.4+mod & Tek1024 \nl
623.66 & 630.66 & 7  &MDM 2.4+mod & SiTe2048 \nl
800.06 & 801.06 & 2  &MDM 2.4+mod & SiTe2048 \nl
842.04 & 846.98 & 8  &MDM 2.4+mod & SiTe2048 \nl
894.90 & 895.98 & 4  &MDM 2.4+mod & SiTe2048 \nl
955.83 & 955.91 & 8 & Lick 3m + Kast & Reticon $1200\times 400$\nl
987.76 & 987.84 & 8 & Lick 3m + Kast & Reticon $1200\times 400$\nl
\enddata
\tablenotetext{a}{Heliocentric Julian date minus 2,450,000.}
\label{tab1}
\end{deluxetable}

All of these observations strongly suggested that PG 1224+309 is a
close binary system, in which the M star is irradiated by the white
dwarf.  We therefore obtained additional photometric and spectroscopic
data on this object, reported in this paper.  Our goals were to
characterize the orbit, to estimate the masses and other properties of
the two stars, and to place the system in its evolutionary context.
The additional data confirm that PG 1224+309 is a short-period,
double-lined spectroscopic binary system with photometric and
emission-line variations evidently caused by illumination of one
hemisphere of the cool M star by the hot white dwarf.  We present
below a description of our observations and data analysis.  We discuss
various constraints on the component masses and the geometry of the
system and end with a brief discussion of the evolutionary status of
this binary star.  Systems such as PG 1224+309 may illuminate
difficult aspects of common-envelope binary evolution.

\section{Observations}

\subsection{MDM\protect\footnote{MDM Observatory is operated by a consortium of
Dartmouth College, Columbia University, the Ohio State University, and
the University of Michigan.} Spectroscopic Observations}

We obtained 55 exposures between 1996 December and 1998 March using
either the 2.4 m Hiltner Telescope and modular spectrograph, or the
1.3 m McGraw-Hill Telescope and Mark III spectrograph.  Table
\ref{tab1} summarizes the observations.  The 2.4 m data had a spectral
resolution of $\sim 3.5$ \AA\ FWHM, and the 1.3 m data had $\sim 5$
\AA , as determined from comparison lines.  All spectra covered
H$\beta$ to H$\alpha$, and the later 2.4 m spectra (taken with the
$2048\times 2048$ CCD) covered from 4000 to 7500 \AA\ with
considerable vignetting toward the ends.  Individual exposures were
300 to 600 s at the 2.4 m telescope, and typically 900 s at the 1.3 m.
All stellar spectra were bracketed in time with comparison-lamp
exposures, and typical rms residuals of the wavelength solutions were
$< 0.1$ \AA.  Standard IRAF\footnote{IRAF is distributed by the
National Optical Astronomy Observatories.}  {\it specred} tasks were
used to subtract the electronic bias, divide by flat-fields, optimally
extract one-dimensional spectra, subtract sky background, and rebin
the spectra onto a uniform wavelength scale.  We obtained spectra of
hot stars and flux standards to convert our spectra to absolute flux
units.


At the 2.4 m we used a narrow slit (1 arcsec), so our
spectrophotometry suffers from variable, uncalibrated losses at the
slit jaws; we estimate that this effect introduces a $\sim 30$ percent
uncertainty in our absolute flux levels.  In addition, continuum
shapes derived with the modular spectrograph sometimes suffer 10--20
per cent errors in their overall slopes, which we do not
understand. We did rotate the spectrograph to align the slit with the
parallactic angle whenever the zenith distance was substantial, so
differential refraction is not the culprit.

\subsection{Kitt Peak Photometric Observations}

We obtained images of PG 1224+309 during the interval 1998 May 5-13
with the KPNO 0.9m telescope, which was equipped with the T2KA
$2048\times 2048$ CCD at both the f/13.5 focus (May 5-8) and the f/7
focus (May 9-13).  Typical exposure times were 400 s for the $B$
filter, 300 s for $V$, 200 s for (Kron-Cousins) $R$, and 250 s for
(Kron-Cousins) $I$.  The night of May 4/5 was photometric and we
obtained 24 observations of 16 different standard stars from the list
of Landolt (1992\markcite{landolt92}) in each of the $B$, $V$, $R$,
and $I$ filters, in addition to the nine $B$, $V$, $R$, and $I$
sequences on PG 1224+309.

\begin{deluxetable}{crcrcrcr}
\tablewidth{0pt}
\tablecolumns{8}
\tablecaption{H$\alpha$ Radial Velocities}
\tablehead{
\colhead{HJD\tablenotemark{a}} & \colhead{V} & 
\colhead{HJD\tablenotemark{a}} & \colhead{V} & 
\colhead{HJD\tablenotemark{a}} & \colhead{V} & 
\colhead{HJD\tablenotemark{a}} & \colhead{V} \\
\colhead{} & \colhead{(km s$^{-1}$)} & 
\colhead{} & \colhead{(km s$^{-1}$)} &
\colhead{} & \colhead{(km s$^{-1}$)} &
\colhead{} & \colhead{(km s$^{-1}$)}
}
\startdata
 $447.998$ &  $ -48$   &  $490.777$ &  $ -73$   &  $624.656$ &  $  57$   &  $846.017$ &  $ 163$   \nl 
 $451.052$ &  $ 125$   &  $491.953$ &  $   2$   &  $624.660$ &  $  47$   &  $846.025$ &  $ 161$   \nl 
 $451.058$ &  $ 120$   &  $491.964$ &  $ -32$   &  $626.658$ &  $ 130$   &  $846.962$ &  $-113$   \nl 
 $487.032$ &  $  20$   &  $497.888$ &  $  86$   &  $626.662$ &  $ 172$   &  $846.969$ &  $-129$   \nl 
 $487.833$ &  $-112$   &  $497.894$ &  $  77$   &  $630.658$ &  $-156$   &  $895.968$ &  $ 144$   \nl 
 $488.859$ &  $   0$   &  $498.894$ &  $ 187$   &  $800.062$ &  $  14$   &  $895.976$ &  $ 138$   \nl 
 $488.870$ &  $   9$   &  $498.900$ &  $ 148$   &  $842.038$ &  $-154$   &  
\nodata \nl
 $489.065$ &  $ 209$   &  $623.658$ &  $ -75$   &  $845.010$ &  $ 173$   &  
\nodata \nl
 $489.840$ &  $ 215$   &  $623.665$ &  $ -97$   &  $846.010$ &  $ 147$   &  
\nodata \nl 
\enddata
\tablenotetext{a}{Heliocentric JD of mid-integration minus 2450000.}
\label{tab2}
\end{deluxetable}

Standard IRAF tasks were used to subtract the electronic bias and
perform flat-field corrections.  We used the programs DAOPHOT IIe,
ALLSTAR and DAOMASTER (Stetson 1987\markcite{st87}; Stetson, Davis, \&
Crabtree 1991\markcite{sdc91}; Stetson 1992a\markcite{st92a},b) to
compute the photometric time series of PG 1224+309 and five nearby
field stars.  The observations of the Landolt standard stars were used
to derive the transformation from DAOMASTER instrumental magnitudes to
the standard scales.  The formal errors for the zeropoints of the
calibrated magnitude scales are $0.018$ mag for the $B$ filter,
$0.005$ mag for the $V$ filter, $0.012$ mag for the $R$ filter, and
$0.013$ mag for the $I$ filter.  Another indication of the quality of
the photometric calibration comes from an examination of the measured
comparison star magnitudes.  Each comparison star had nine measures in
the four different filters.  There was very little scatter in these
measurements: the standard deviations ranged from 0.003 to 0.030 mag.

\subsection{Lick Spectroscopic Observations}

Blue spectra of PG~$1224+309$ were obtained with the 3m Shane
Telescope and Kast spectrograph (Miller \& Stone 1993\markcite{ms93})
at Lick Observatory on 1998 May 22 UT and again on 1998 June 23 UT. On
each occasion the star was observed for approximately 2 hours during
which a set of eight 15-minute exposures was obtained, covering about
1/3 of the orbital cycle. The purpose was to measure the radial
velocity curve of the white dwarf from the high-order Balmer lines in
its spectrum, so the observations were planned to sample a different
part of the orbital cycle of the system on each run. The spectra cover
the range 3200--4550~\AA\ with a spectral resolution of 2.8~\AA.  They
were obtained through a $2^{\prime\prime}$ slit using an 830~mm$^{-1}$
grism. The seeing varied between $1^{\prime\prime}\!\!.5$ and
$2^{\prime\prime}$ over the course of the observations resulting in a
variable loss of light at the slit jaws. Moreover, since the star had
to be followed through a fairly high airmass, some spectra suffer from
additional light losses at the shortest wavelengths because of
differential atmospheric refraction, in spite of our efforts to keep
the slit at the parallactic angle.  A log of the observations is
included in Table \ref{tab1}.

The spectra were reduced in a standard manner, as described earlier,
although we used our own reduction software while making only limited
use of the IRAF software package. In the final individual spectra, the
signal-to-noise ratio at 4200~\AA\ ranges between 15 and 25. The
wavelength scale was originally derived from spectra of arc lamps
obtained at the beginning of each sequence of object spectra. The
wavelength scales of individual spectra were then refined by using the
emission-line spectra of the night sky recorded during the same
exposure to measure small {\it relative} offsets and rectifying
them. As a result of this procedure, the {\it relative} velocity
scales of individual spectra agree to 8~km~s$^{-1}$ or better.

\section{Data Analysis}

\subsection{The Spectroscopic Period and the H$\alpha$ Emission Line
Radial Velocity Curve}

\begin{figure}[t]
\vspace{2in}
\includegraphics{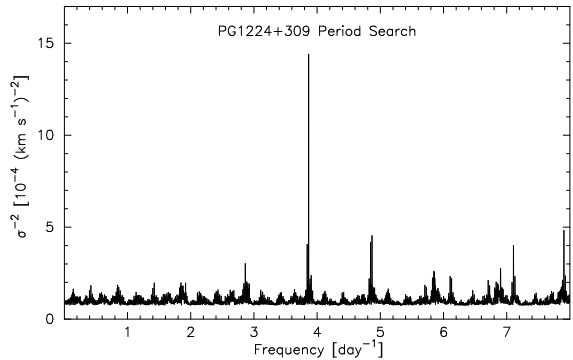}
\caption{Results of a period search on the H$\alpha$ emission
velocities.  The vertical axis is the inverse of the variance 
around a sinusoidal fit at each trial frequency.
In order to compress the large number of data points we show 
local maxima in this function, joined by straight lines.  
}
\label{t1}
\end{figure}

\begin{figure}[h]
\vspace{4.1528in}
\includegraphics{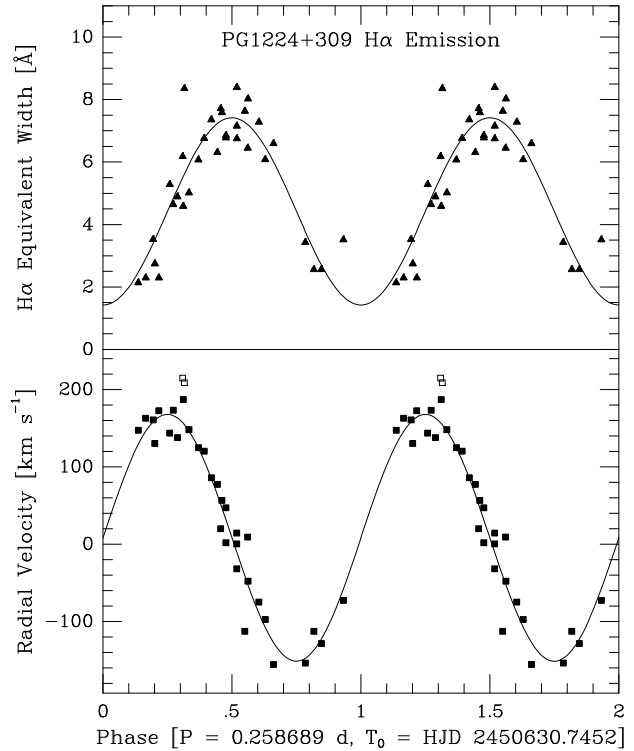}
\caption{Lower panel: radial velocities of the H$\alpha$ emission lines, 
derived from Gaussian fits to the line profile, folded on 
the best period, with the best-fitting sinusoid superposed.
The points represented by open squares were excluded from the 
fit. Upper panel: Equivalent widths of the H$\alpha$ emission
line folded on the best period.  The fiducial sinusoid
shown is shifted by 0.25 cycle from the velocity ephemeris.
}  
\label{t2}
\end{figure}

We used the set of spectra from MDM to determine the spectroscopic
period since the time coverage for these data is by far the longest.
The MDM spectra (discussed later) showed a narrow, variable H$\alpha$
emission line superposed on a strong white dwarf spectrum of type DA
(see also Orosz, Wade, \& Harlow 1997\markcite{orosz1}).  For 33
spectra with adequate signal-to-noise, we fitted this feature with a
Gaussian, adjusting approximately for the sloping baseline caused by
the white dwarf absorption.  Table \ref{tab2} gives the resulting
radial velocities.  A Gaussian-convolution algorithm (Schneider \&
Young 1980\markcite{sy80}) gave similar results, but with slightly
greater scatter around the best fit (below).  We searched for periods
by constructing a highly oversampled grid of trial frequencies
spanning 0 to 8 cycles d$^{-1}$, fitting a least-squares sinusoid to
the velocities at each frequency, and plotting the inverse of the fit
variance as a function of frequency to form a `residual-gram'
(Thorstensen et al.~1996\markcite{tpst}).  This method works well when
the modulation is sinusoidal and its amplitude exceeds the
observational uncertainty.  Figure \ref{t1} shows the result, which is
consistent with a single frequency near 3.86 d$^{-1}$; the cycle count
is unambiguous.  A least-squares sinusoid fit of the form
$$
v(t) = \gamma_{{\rm H}\alpha} + 
K_{{\rm H}\alpha} \sin\left[2 \pi (t - T_0) \over P\right]
$$
gave
$$
\begin{array}{rcl}
T_0 & = & \hbox{HJD}\ 2450624.537 \pm 0.002, \\
P & = & 0.258689 \pm 0.000004 \ {\rm d}, \\
K_{{\rm H}\alpha} & = &  165 \pm 9\ {\rm km\ s^{-1}}, \\
\gamma_{{\rm H}\alpha} & = & 10 \pm 6\ {\rm km\ s^{-1}},\ \hbox{and} \\
\sigma & = & 29\ {\rm km\ s^{-1}},
\end{array}
$$
where $\sigma$ is the uncertainty of a single measurement inferred
from the goodness-of-fit, and the uncertainties are formal 1-$\sigma$
errors.  Discarding the two worst-fitting points changed $K$,
$\gamma$, and $\sigma$ to $160 \pm 8$, $8 \pm 5$, and $26$ km s$^{-1}$
respectively, but did not significantly affect the ephemeris.  Figure
\ref{t2} (lower panel) shows all the velocities folded on the
ephemeris.  The sinusoid shown has been fitted to all but the two
worst points.  We adopt $K_{{\rm H}\alpha}=160\pm 8$ km~s$^{-1}$.


The H$\alpha$ emission fits also produced estimates of the equivalent
width (EW).  These were noisier than the radial velocities, but showed
a strong modulation at the same period.  A sinusoidal fit to the EW
time series, with the period fixed at the more accurately-determined
radial velocity period, showed that the EW modulation lags in phase
behind the radial velocities by $0.256 \pm 0.016$ cycle.  This is
nicely consistent with the 0.25-cycle lag expected if the emission
line arises on the side of the normal star facing the white dwarf
(Thorstensen et al.~1978\markcite{tcmb78}; Vennes \& Thorstensen
1994\markcite{vt94}).  The upper panel of Figure \ref{t2} shows the
equivalent widths folded on the best period, together with a
sinusoidal fit with a half-amplitude of $3.0 \pm 0.3$ \AA\ and a mean
of $4.4 \pm 0.2$ \AA\ (formal error).  The sinusoidal fit is only
fiducial, as the modulation may not be strictly sinusoidal
(Thorstensen et al.~1978\markcite{tcmb78}), and our EW accuracy and
sampling are not sufficient for us to detect small deviations from a
sinusoid.  In any case, the deep modulation of the equivalent widths,
and the accurate agreement of the phase offset with the illumination
model, make it very likely that the bulk of the H$\alpha$ emission
arises from fluorescence on the illuminated face of the normal star.
It is possible that the M dwarf also has intrinsic H$\alpha$ emission,
but it apparently does not dominate.

\begin{deluxetable}{ccccccc}
\tablewidth{0pt}
\tablecolumns{7}
\tablecaption{Photometric parameters}
\tablehead{
\colhead{Filter} & \colhead{$N$}  &\colhead{Period} & 
\colhead{Semi-amplitude} 
& \colhead{Mean} & \colhead{$T_0({\rm photo})$}
& \colhead{Spectroscopic phase} \\
\colhead{} &\colhead{} & \colhead{(days)} & 
\colhead{(mag)} 
& \colhead{} & \colhead{(HJD 2450000+)}
& \colhead{of $T_0({\rm photo})$}}
\startdata
$B$ & 38  &  $0.2587 \pm 0.0003$  &  $0.034 \pm 0.002$  &  $16.099 \pm 0.002$
                               &  $938.840\pm 0.004$ & $1214.98\pm 0.03$ \nl
$V$ & 55  &  $0.2588 \pm 0.0001$  &  $0.070 \pm 0.002$  &  $16.164 \pm 0.001$
                               &  $938.845\pm 0.002$ & $1215.00\pm 0.02$ \nl
$R$ & 39  &  $0.2586 \pm 0.0001$  &  $0.104 \pm 0.003$  &  $16.126 \pm 0.002$
                               &  $938.847\pm 0.002$ & $1215.01\pm 0.02$ \nl
$I$ & 50  &  $0.2582 \pm 0.0001$  &  $0.098 \pm 0.002$  &  $15.812 \pm 0.001$
                               &  $938.849\pm 0.001$ & $1215.02\pm 0.02$ \nl
\enddata
\label{tab3}
\end{deluxetable}

\begin{figure}[p]
\vspace{4.1528in}
\includegraphics{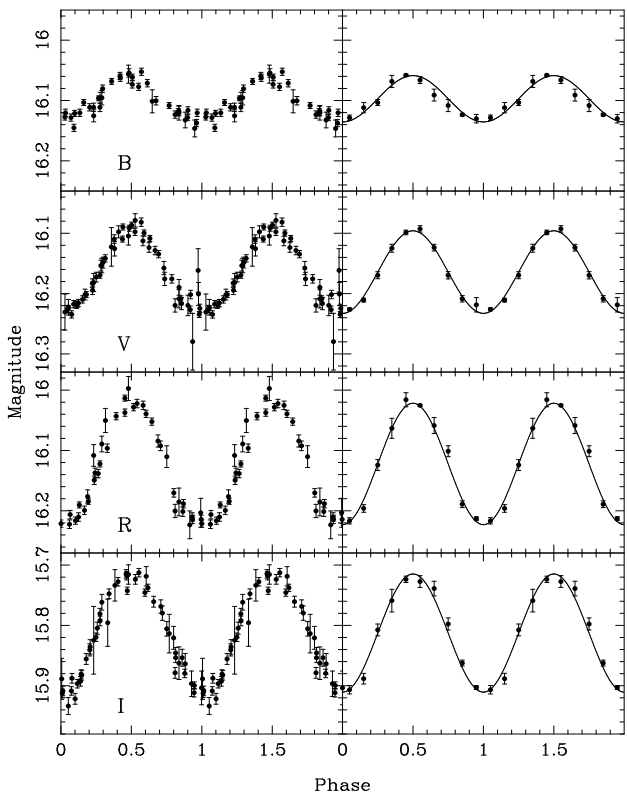}
\caption{The $B$, $V$, $R$, and $I$ light curves folded on the
adopted (mean) ephemeris are shown in the left panels, where each point
has been plotted twice.  The right panels show the folded light curves
binned into ten phase bins, where the error bars represent the errors
of the mean in each bin.  Sine curves have been added to demonstrate
the absence of eclipses; the minimum of each curve occurs at phase
zero, and the semiamplitudes in the $B$, $V$, $R$, and $I$ bands are 0.04,
0.07, 0.10, and 0.10 mag respectively.  }
\label{o5}
\end{figure}

\subsection{Photometric Variations}

The $B$, $V$, $R$, and $I$ light curves of PG 1224+304 display
approximately sinusoidal modulations with a period near $\approx 6$
hours, as expected based on the spectroscopic analysis.  We fit
four-parameter sinusoids to each of the four light curves (using the
spectroscopic period $P_{\rm spect}$ as the initial period guess) to
determine the photometric period, semi-amplitude, mean magnitude, and
the time of photometric minimum.  The results are displayed in Table
\ref{tab3}.  The errors on the individual magnitudes were scaled to
give $\chi^2_{\nu}=1$ for each curve and the errors on the fitted
parameters were computed from the scaled errors.  The errors listed in
Table \ref{tab3} may have been slightly underestimated, since the light
curve shapes may not be exact sinusoids.  The periods derived for the
four light curves are close to the spectroscopic period.  We also
computed the spectroscopic phase of the time of the photometric
minimum for each light curve.  All of these phases are consistent with
a whole number of cycles, as expected if the cause of the modulation
is due to irradiation of the M star by the white dwarf.  We used the
weighted mean of the photometric $T_0$ and the spectroscopic $T_0$
(propagated forward by 1207 cycles) to define the phase zero-point for
the adopted ephemeris: $T_0({\rm mean})={\rm HJD}\, 2,450,938.8430\pm
0.0026$.  Our adopted orbital parameters are given in Table
\ref{tab4}, and the folded light curves are displayed in Figure
\ref{o5}.

\begin{deluxetable}{rr}
\tablewidth{0pt}
\tablecolumns{2}
\tablecaption{Adopted Orbital Parameters}
\tablehead{
\colhead{parameters}  &
\colhead{value}}
\startdata
$P_{\rm spect}$ (days) &  $0.258689\pm 0.000004$  \nl
$K_{{\rm H}\alpha}$ (km s$^{-1}$)  &  $160 \pm 8$  \nl
$K_{\rm WD}$ (km s$^{-1}$)  &  $112 \pm 14$  \nl
$T_0$(spect) (HJD 2,450,000+)  &  $626.608\pm 0.002$ \nl
$T_0$(photo) (HJD 2,450,000+)  &  $938.844\pm 0.003$ \nl
$T_0$(mean) (HJD 2,450,000+)  &  $938.843\pm 0.003$ \nl
\enddata
\tablecomments{$T_0$ refers to the sine-fit of radial velocities
measured from the H$\alpha$ emission line, i.e.\ inferior conjunction
of the M star.}
\label{tab4}
\end{deluxetable}

\subsection{The White Dwarf Radial Velocity Curve}
%
%

\begin{figure}[h]
\vspace{1.55in}
\includegraphics{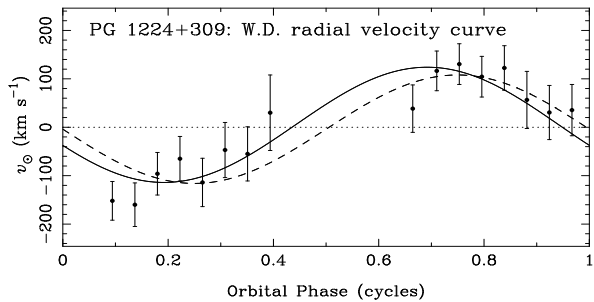}
\caption{The measured radial velocity curve of the white dwarf with
two circular-orbit models superposed for comparison. The radial
velocities were measured from the high-order Balmer lines (H$\gamma$,
H$\delta$, and H$\epsilon$) and have been corrected for the Earth's
motion around the Sun. The orbital phase is computed according to the
adopted (mean) ephemeris. The solid line shows a fit in which the phase
of inferior conjunction of the white dwarf was treated as a free parameter,
while the dashed line shows a model in which this phase was fixed at
0.5.}
\label{mike_new}
\end{figure}

The radial velocity of the white dwarf was measured from each Lick
spectrum by comparing it to a synthetic white dwarf spectrum,
constructed with the parameters reported by Liebert \& Bergeron
(1997\markcite{lieb}).  The method involves assuming a radial velocity
for the white dwarf, applying a Doppler shift and appropriate scaling
to the model spectrum, and comparing it to the observed spectrum in
the range 3890--4400~\AA\ by means of the $\chi^2$ test. By scanning a
range of possible white dwarf radial velocities in very fine steps we
obtain the best estimate as the one with the lowest value of $\chi^2$
as well as error bars corresponding to 68\% (1-$\sigma$) confidence
limits. The measured heliocentric radial velocities are given in Table
\ref{mikeobs} along with their corresponding orbital phases, computed
from the adopted mean ephemeris given in Table \ref{tab4}.  In Figure
\ref{mike_new} we plot the radial velocity of the white dwarf as a
function of this orbital phase.


The observed radial velocities were fitted with a circular-orbit
model.  Because the observations do not cover a complete orbital cycle
we cannot use them to obtain a measurement of the period. Hence we
adopt the period derived from the H$\alpha$ velocities and we express
the model as
\begin{equation}
v(\phi)=\gamma_{\rm WD} + K_{\rm WD}\; \sin\left[2\pi(\phi-\Delta\phi)\right] ,
\end{equation}
where $\phi$ is the orbital phase from the adopted ephemeris
($0\leq\phi<1$) and $\Delta\phi$ is the phase of inferior conjunction
of the white dwarf, which is expected {\it a priori} to be 0.5. The
free parameters of the model are $\gamma_{\rm WD}$, $K_{\rm WD}$ and
$\Delta\phi$. A minimum $\chi^2$ fit of the model to the data gives
the following parameters:
$$
\begin{array}{rcl}
\Delta\phi & = & 0.442^{+0.049}_{-0.044}, \\
K_{\rm WD} & = &  119^{+24}_{-23} ~{\rm km\ s^{-1}},\ {\rm and}\\
\gamma_{\rm WD} & = & 5 \pm 21\ {\rm km\ s^{-1}} \\
\end{array}
$$
with errors corresponding to 68\% confidence limits. Since the
best-fitting value of $\Delta\phi$ is consistent with expectation
(according to the F-test), we may fix $\Delta\phi$ to its expected
value. Under this assumption we obtain:
$$
\begin{array}{rcl}
\Delta\phi & = & 0.5~~{\rm (fixed)}, \\
K_{\rm WD} & = &  112 \pm 14\ {\rm km\ s^{-1}},\ {\rm and}\\
\gamma_{\rm WD} & = & -4 \pm 12\ {\rm km\ s^{-1}} \\
\end{array}
$$
The best-fitting model radial velocity curves (with $\Delta\phi$ both
free and fixed) are superposed on the data in Figure
\ref{mike_new}, for comparison.

\begin{deluxetable}{rrrrrrr}
\tablewidth{0pt}
\tablecolumns{7}
\tablecaption{Heliocentric Radial Velocities from White Dwarf Absorption Lines}
\tablehead{
\colhead{HJD\tablenotemark{a}} & \colhead{$\phi$\tablenotemark{b}} & \colhead{$V
_{\odot}$} &
\colhead{} &
\colhead{HJD\tablenotemark{a}} & \colhead{$\phi$\tablenotemark{b}} & \colhead{$V
_{\odot}$} \\
\colhead{} & \colhead{} & \colhead{(km s$^{-1}$)} &
\colhead{} &
\colhead{} & \colhead{} & \colhead{(km s$^{-1}$)}
}
\startdata
 955.82971 & 0.665 & $  38 \pm 49$ &\phantom{space}&  987.75956 & 0.094 & $-152 
\pm 40$ \nl    
 955.84151 & 0.710 & $ 116 \pm 41$ & &  987.77064 & 0.137 & $-160 \pm 45$ \nl   
    
 955.85258 & 0.753 & $ 130 \pm 42$ & &  987.78170 & 0.180 & $ -96 \pm 44$ \nl   
    
 955.86365 & 0.796 & $ 104 \pm 42$ & &  987.79276 & 0.222 & $ -65 \pm 46$ \nl   
    
 955.87471 & 0.839 & $ 122 \pm 46$ & &  987.80385 & 0.265 & $-114 \pm 50$ \nl   
    
 955.88578 & 0.881 & $  56 \pm 59$ & &  987.81493 & 0.308 & $ -47 \pm 57$ \nl   
    
 955.89683 & 0.924 & $  30 \pm 56$ & &  987.82601 & 0.351 & $ -55 \pm 56$ \nl   
    
 955.90792 & 0.967 & $  35 \pm 53$ & &  987.83709 & 0.394 & $  30 \pm 78$ \nl  \enddata
\tablenotetext{a}{Heliocentric JD of mid-integration minus 2450000.}
\tablenotetext{b}{Orbital phase, computed from the adopted (mean) ephemeris.}
\label{mikeobs}
\end{deluxetable}

\subsection{White Dwarf Atmospheric Parameters}

\begin{figure}[h]
\includegraphics{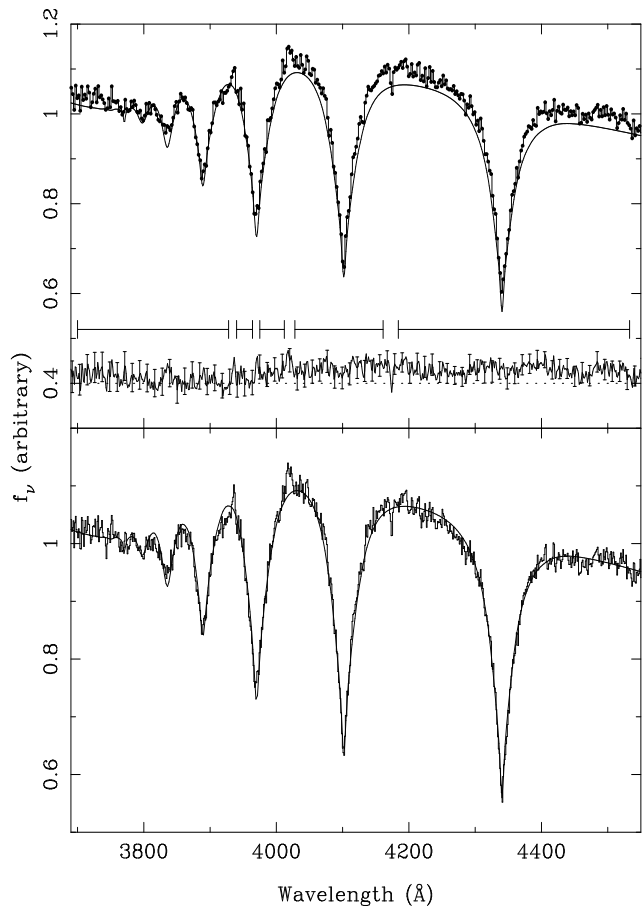}
\caption{Upper panel, from top to bottom: 
The blue ``restframe'' spectrum of PG 1224+309, constructed from four
spectra having a mean orbital phase of $\phi=0.16$ (filled circles);
the best-fitting model spectrum (with $\log g=7.38$ and $T_{\rm
eff}=29,300$ K), scaled by 0.967 (smooth line); and the difference
spectrum with error bars on the individual points, offset upwards by
0.4 for clarity.  The horizontal bars with gaps indicate the
wavelength regions used in the fit.  Lower panel: The best-fitting
model (smooth line) is shown with the detrended data, made by
subtracting the polynomial fit to the residuals from the observed
spectrum.}
\label{newspectfit}
\end{figure}

Using the fitted radial velocity curve of the white dwarf, we
corrected the Doppler shifts in the Lick spectra and averaged them
together to obtain a spectrum with a high signal-to-noise ratio, which
is free from orbital Doppler broadening. We chose to use four of the
Lick spectra for this purpose, namely the ones spanning orbital phases
0.09 to 0.22. These four spectra were obtained under the best seeing
conditions and at the lowest airmass. As a result they individually
have a high signal-to-noise ratio and the effects of differential
atmospheric refraction are negligible. Also, the line-emitting face of
the secondary star is directed mainly away from the observer at these
orbital phases. Hence the contamination of the cores of the absorption
lines in the white dwarf spectrum by narrow emission lines is
minimal. The average spectrum of the white dwarf is shown in Figure
\ref{newspectfit}.


We used this Lick ``restframe'' spectrum with its improved
signal-to-noise to estimate values of the white dwarf surface gravity
$\log g$ and effective temperature $T_{\rm eff}$.  
(Recall that
$T_{\rm eff}=29,300$ K and $\log g=7.38\pm 0.1$ according to Liebert
\& Bergeron [1997]\markcite{lieb}.  Their uncertainties include
estimates of the systematic errors due to modeling assumptions and
input physics.)  We computed a grid of pure hydrogen, LTE model
spectra, covering the range $\log g=7.18$ to 7.58 in steps of 0.1 and
the range $T_{\rm eff}=28,300$ K to 29,800 K in steps of 500 K.  These
synthetic spectra were computed using I. Hubeny's programs TLUSTY and
SYNSPEC, and the broadening of the hydrogen lines was treated using
the tables of Schoening and Butler (see Hubeny \& Lanz
1997\markcite{hubeny}).  We fitted the Lick restframe spectrum over
the wavelength interval 3700--4533~\AA, excluding the Ca II H and K
lines and features at approximately $4018$ and 4174~\AA.

The observed spectrum of PG 1224+309 consists of {\em three}
components: the white dwarf spectrum, the spectrum of the late-type
secondary (see below), {\em and} the spectrum from the heated face of
the secondary.  It is likely that the unilluminated portion of the M
star does not make a noticeable contribution in the $U$ and $B$
bands, but the same thing cannot be said about the heated face of the
M star.  The Lick restframe spectrum was constructed so that the
contamination from the heated face was minimized as much as possible,
but the fact that the H$\alpha$ emission line is always visible
suggests that some of the heated hemisphere is visible near phase 0.0.
To allow for this extra light, we used a fitting method similar to the
one given in 
Marsh, Robinson, \& Wood (1994\markcite{marsh}), 
which can be summarized
as follows.  The data and models are all normalized at a common
wavelength (4501 \AA).  Each model spectrum is scaled by an amount $w$
($0\le w\le 1$) and subtracted from the data, a third-order polynomial
is fitted to the residuals, and the rms residual of the fit is
recorded.  The optimal fit is defined to be the one that gives the
lowest overall value for the rms, corresponding to the ``smoothest''
difference spectrum.  Here we are assuming that the spectrum of the
irradiated hemisphere is featureless.  The best decomposition is shown
in Figure \ref{newspectfit}.  The model white dwarf spectrum has
$T_{\rm eff}=29,300$ K and $\log g=7.38$ and contributes 97\% of the
light at 4500~\AA\ and virtually all of the light blueward of
3800~\AA.  Our values of $T_{\rm eff}$ and $\log g$ are the same
within the errors as those found by 
Liebert \& Bergeron\markcite{lieb}. 
Thus we adopt
$\log g=7.38$ for the discussion below.

\vfill

\newpage

\subsection{The Spectral Type of the Secondary Star}

\begin{figure}[t]
\vspace{3.9in}
\includegraphics{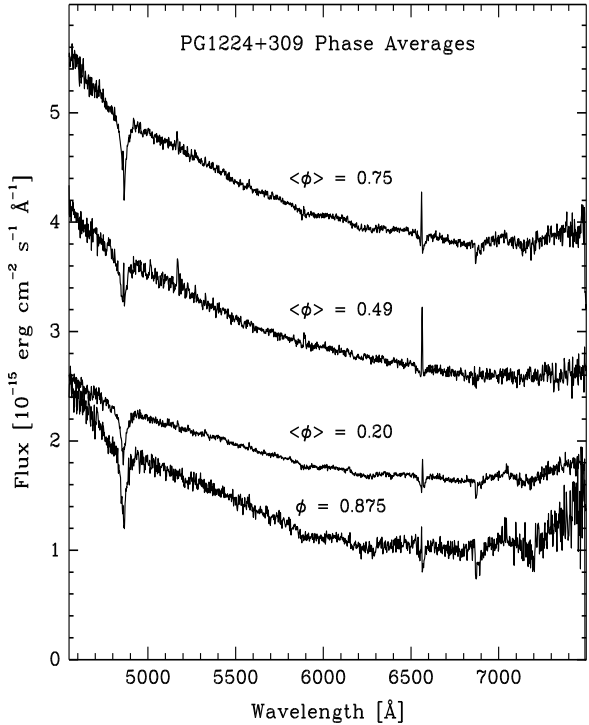}
\caption{Average spectra of PG 1224+309 in 
0.25-cycle phase bins. Each is labeled by its mean phase.
Only one spectrum fell in the bin nearest zero phase.
No scale factors are applied, and the spectra are offset
from each other vertically 
by one unit ($10^{-15}$ erg cm$^{-2}$ s$^{-1}$
\AA $^{-1}$).
}
\label{t3}
\end{figure}

Having knowledge of the orbital ephemeris, we searched in the red
spectra for the contribution of the M star and for any variations of
its spectrum with orbital phase.  Figure \ref{t3} shows our
large-format MDM 2.4 m spectra (which cover to $\lambda = 7500$ \AA)
averaged into $1/4$-cycle phase bins centered on the four cardinal
phases (conjunction and quadrature).  Unfortunately, only a single
spectrum fell in the bin nearest inferior conjunction of the M star.
Nonetheless, a heating effect is clearly seen.  In the lowermost
spectrum (inferior conjunction), the characteristic bumpy continuum of
an M dwarf is strongly present; it is less pronounced, though quite
obvious, in the quadrature phases.  However, in the spectrum showing
the largest portion of the illuminated face (third from bottom), the
M-dwarf contribution is not obvious despite good signal-to-noise.
Evidently the spectrum of the illuminated face is of earlier type than
the unilluminated face (or is otherwise lacking strong molecular
bands), so that the bands in the observed spectrum are greatly
diluted.  The illumination of the M star is discussed further below.


\begin{figure}[t]
\vspace{2.4in}
\includegraphics{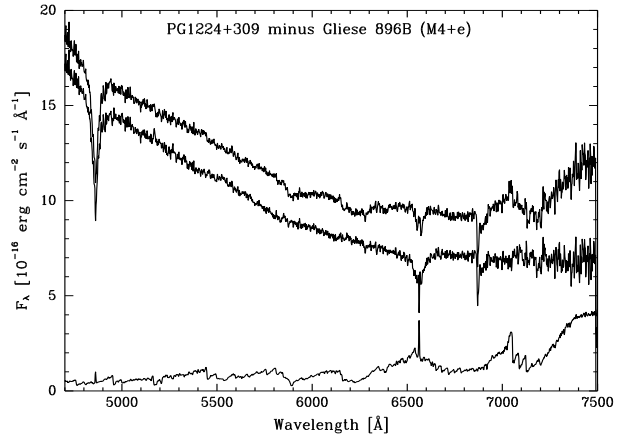}
\caption{Decomposition of the spectrum.  The top trace is the
average of eight spectra taken within 0.2 cycles of inferior conjunction
of the M dwarf.  The lower trace is a scaled spectrum of Gliese 896B,
spectral type M4+ 
(Boeshaar 1976) or M4.5V (Henry, Kirkpatrick, \& Simons 1994).
The middle trace is the difference of the two.  The top trace is
offset upward by one unit ($10^{-15}$ erg cm$^{-2}$ s$^{-1}$ \AA $^{-1}$)
to avoid overlap with the middle trace.
}
\label{t4}
\end{figure}

To quantify the spectral contribution of the secondary star and
estimate its type, we prepared an average of the eight large-format
spectra taken within 0.2 cycle of inferior conjunction.  We then
systematically scaled and subtracted spectra from a library of M
dwarfs.  These stars were observed with the same instrumental setup,
and have classifications from Boeshaar (1976)\markcite{boes}.  We
examined the subtracted spectra looking for good cancellation of the
M-dwarf features.  Types M3 and earlier tended to leave a strong
oversubtraction of the \ion{Na}{1} D lines when the scale factor was
set to cancel the bands in the 6500--7500 \AA\ region, and the bands
themselves did not cancel well.  A spectrum of type M5 did not
subtract satisfactorily.  The only acceptable results were from two
stars classified as M4+ (Boeshaar 1976\markcite{boes}) or M4.5 (Henry,
Kirkpatrick, \& Simons 1994\markcite{hen1}).  Figure \ref{t4} shows the
best decomposition.  We conclude that the intrinsic spectrum of the M
dwarf is most probably a little later than M4, and that
the flux contributed at this phase by the unilluminated part of the M
dwarf is $F_{\lambda} \sim 1.5 \times 10^{-16}$ erg cm$^{-2}$ s$^{-1}$
\AA $^{-1}$ at 6500 \AA, or 15--20 percent of the light at that
wavelength.  The calibration uncertainties discussed earlier should
not seriously affect these conclusions, since the features cover a
fairly limited wavelength range.  For comparison, Orosz, Wade, \&
Harlow (1997\markcite{orosz1}) adopted a spectral type of M1--M3 with
a contribution of 10\% over the region 5320--6532~\AA, based on two
low signal-to-noise spectra obtained near orbital phases 0.89 and
0.53.


Very few features are seen in the spectrum aside from the
contributions discussed above.  There is weak emission present at
$\lambda 5169$, apparently also modulated by the illumination phase.
This feature is fairly often seen in cataclysmic binaries, where it is
usually attributed to Fe II (Taylor \& Thorstensen
1996\markcite{tt96}).  Curiously, there is very little He I emission;
He I is conspicuous (though weaker than the Balmer lines) in the
illumination-effect stars Feige 24 (Thorstensen et
al.~1978\markcite{tcmb}) and EUVE 2013+400 (Thorstensen et
al.~1994\markcite{tbs94}).

Because the H$\alpha$ emission line arises mostly from one hemisphere
of the M dwarf, its velocity amplitude is likely to underestimate the
star's center-of-mass motion.  To account for this effect we tried to
measure velocities of the M dwarf by cross-correlation methods, using
our M dwarf library spectra (some of which have very accurate
velocities from Marcy et al. 1987\markcite{mlw87}) as templates.
Unfortunately, the signal-to-noise ratio of the available spectra
proved inadequate for this purpose.

\section{Discussion}

\subsection{Mass Constraints}

Knowing the surface gravity of the white dwarf in PG 1224+309 enables
us to estimate its mass.  The mass--radius relation for white dwarf
stars does depend on the composition of the white dwarf and its
position along the cooling track.  In addition, the radius will be
larger for a white dwarf that has a hydrogen layer atop the degenerate
core.  
Liebert \& Bergeron (1997\markcite{lieb}) 
estimated the mass of PG 1224+309 to be
$M_{\rm WD}=0.36\,M_{\sun}$ based on their estimates of $T_{\rm
eff}=29,300$ and $\log g=7.38$ and using then-available evolutionary
tracks for carbon-oxygen white dwarfs 
(cf.\ Wood \& Winget 1989\markcite{wood}).  For
these same values of $\log g$ and $T_{\rm eff}$, evolutionary tracks
for helium white dwarfs by 
Althaus \& Benvenuto (1997)\markcite{alt} 
yield $M_{\rm
WD}=0.40\,M_{\sun}$ (their Fig.\ 13).  These models do not include a
hydrogen envelope.  
Benvenuto \& Althaus (1998)\markcite{benalt} 
published additional
cooling tracks for helium white dwarfs that include a thin hydrogen
envelope.  At $\log g = 7.38$, these models imply $0.40\le M_{\rm
WD}\le 0.45\,M_{\sun}$ for hydrogen envelope masses of $0\le M_{\rm H}
\le 4\times 10^{-4}\,M_{\odot}$.  For the discussion below we will
adopt $0.35\le M_{\rm WD}\le 0.45\,M_{\sun}$.

\vfill

\begin{figure}[h]
\includegraphics{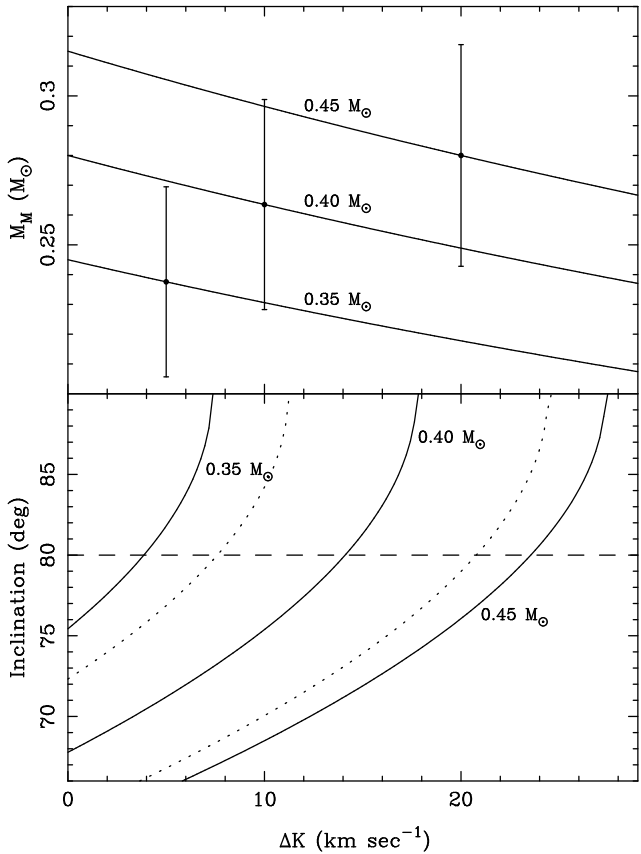}
\caption{Top:  The mass of the M-dwarf secondary star as a function
of $\Delta K$ for $M_{\rm WD}=0.35\,M_{\odot}$,
$M_{\rm WD}=0.40\,M_{\odot}$, and
$M_{\rm WD}=0.45\,M_{\odot}$, assuming $K_{\rm WD}=112\pm 14$ km s$^{-1}$
and $K_{{\rm H}\alpha}=160\pm 8$ km s$^{-1}$.  The sizes of the statistical
errors on $M_{\rm M}$ are illustrated by the three points with error bars.
Bottom:
The derived inclination as a function of $\Delta K$ for
$M_{\rm WD}=0.35\,M_{\odot}$, and
$M_{\rm WD}=0.40\,M_{\odot}$,
$M_{\rm WD}=0.45\,M_{\odot}$.  The dashed lines indicate the range
of allowed solutions for the $M_{\rm WD}=0.40\,M_{\odot}$ case,
propagating one-sigma errors in $K_{\rm WD}$ and $K_{{\rm H}\alpha}$.}
\label{deltaK}
\end{figure}

\newpage

The measured orbital velocities of the two stars determine the mass
ratio: $Q=M_{\rm M}/M_{\rm WD}= K_{\rm WD}/K_{\rm M}$.  If $Q$ were
known accurately, then $M_{\rm M}$ and the total mass of PG 1224+309
would simply scale from $M_{\rm WD}$ as estimated above.  The velocity
semiamplitude of the white dwarf was measured reliably from the higher
Balmer lines visible in the Lick spectra: $K_{\rm WD}= 112\pm 14$ km
s$^{-1}$.  On the other hand, the M star's orbital speed is not
directly measured, since the center of light of the H$\alpha$ line
emission does not correspond to the center of mass of the M star,
$K_{\rm M}=K_{{\rm H}\alpha}+\Delta K$.  Since the H$\alpha$ emission
line is formed in the heated face of the M star, we expect $\Delta K >
0$, or $Q = K_{\rm WD}/(K_{{\rm H}\alpha}+\Delta K) < 0.700\pm 0.094$,
using the adopted $K_{{\rm H}\alpha} =160\pm 8$ km s$^{-1}$.

We can express $\Delta K$ as 
$$
\Delta K = {\Delta R\over a}(1+ Q)K_{\rm M}
$$
where $\Delta R$ is the shift of the ``effective center'' of the
H$\alpha$ emission line with respect to the center of mass of the
M star 
(Wade \& Horne 1988\markcite{wadeh}) and the orbital separation is
$a=1.71(M_{\rm total}/M_{\odot})^{1/3}\,R_{\odot}$.  In the case of a
synchronously rotating M star with uniform H$\alpha$ emission over one
hemisphere and no H$\alpha$ emission over the other, $\Delta R=4R_{\rm
M}/(3\pi)$, where $R_{\rm M}$ is the radius of the M star. To complete
an estimate of $\Delta K$, we temporarily adopt values for $R_{\rm M}$
of $0.25\pm 0.01\,R_{\odot}$ and for $M_{\rm M}$ of $0.25\pm
0.01\,M_{\odot}$, based on the dM4e eclipsing binary CM Draconis 
(Lacy 1977\markcite{lacy}).  
Using $M_{\rm WD}=0.4\pm 0.05\,M_{\odot}$, we then find
$Q=0.625$ and $M_{\rm total}=0.65\,M_{\odot}$.  Using $K_{\rm H\alpha}
=160\pm 8$ km s$^{1}$, we find $\Delta K=21\pm 2$ km s$^{-1}$.

The true value of $\Delta K$ probably cannot be lower than our rough
estimate of 21 km s$^{-1}$, since $\Delta R$ will be larger if the
H$\alpha$ emission is more concentrated to the regions of the M star
closest to the white dwarf (we argued above in Section 3.1 that
the M star probably has very little intrinsic H$\alpha$ emission).  
In principle, $\Delta R$ could be as
large as $R_{\rm M}$ ($\Delta K = 49$~km s$^{-1}$), but then one would
not expect to see the equivalent width of H$\alpha$ varying
sinusoidally, as actually observed.  Tighter upper limits on $Q$ and
$\Delta K$ come from considering the orbital inclination, which must
be less than about $80^{\circ}$ since there are no eclipses longer
than about 0.03 in phase. (Light curve modeling is discussed further
below.)

For an assumed value of $\Delta K$, the measured values of $P$,
$K_{\rm WD}$ and $K_{\rm H\alpha}$, and our estimate of $M_{\rm WD}$
from $\log g$, we compute $Q$ and $M_{\rm total}$, and we infer the
orbital inclination $i$.  This is shown in the lower panel of Figure
\ref{deltaK} for $M_{\rm WD}=0.35\,M_{\sun}$, $0.40\,M_{\sun}$, and
$0.45\,M_{\sun}$.  The upper panel of Figure \ref{deltaK} shows
$M_{\rm M}$ vs.\ $\Delta K$.  If $M_{\rm WD}=0.35\,M_{\odot}$, $\Delta
K$ must be less than $\approx 7.2$ km s$^{-1}$ to allow $i \leq
90^{\circ}$, and if $i< 80^{\circ}$ as suggested by the absence of
eclipses, $\Delta K\lesssim 4$ km s$^{-1}$, much less than our
estimate of $\Delta K\approx 21$ km s$^{-1}$.  For $M_{\rm
WD}=0.4\,M_{\sun}$ and $i<80^{\circ}$, allowed values of $\Delta K$
($\lesssim 14$ km s$^{-1}$) are still smaller than 21 km s$^{-1}$.  In
order to have $\Delta K\gtrsim 21$ km s$^{-1}$, $M_{\rm WD} \gtrsim
0.45\, M_{\sun}$ is required. Taken at face value, this ``large''
white dwarf mass favors the white dwarf models with a ``thick''
hydrogen envelope ($M_{\rm H} = 4\times 10^{-4}\,M_{\sun}$, 
Benvenuto \& Althaus 1998\markcite{benalt}).


For $\Delta K=21\pm 2$ km s$^{-1}$ and $M_{\rm WD}=0.45\pm
0.05\,M_{\odot}$, the mass of the M star is $M_{\rm M}=0.28\pm
0.05\,M_{\odot}$, $i=77\pm 7^{\circ}$, and $Q=0.62\pm 0.08$, where
we have propagated the errors on $K_{\rm WD}$ and $K_{{\rm
H}\alpha}$.  The mass of the M star is not too different from the
measured masses of the dM4e components of CM Draconis: $M=0.237$ and
$0.207\,M_{\odot}$, with errors of 4\% 
(Lacy 1977\markcite{lacy}).  
When $M_{\rm
WD}=0.45\,M_{\odot}$, increasing $\Delta K$ by 5~km~s$^{-1}$ reduces
$M_{\rm M}$ by $0.015\,M_{\odot}$.

To summarize the argument: so far we have discussed a mass estimate
for the secondary star based on orbital dynamics.  The two uncertain
quantities in the analysis are $M_{\rm WD}$, made uncertain by the
unknown H envelope mass, and $\Delta K$.  Adopting a plausible value
for $\Delta K$ (based partly on the radii of the components of CM Dra,
which has a similar spectral type) leads to an estimate of $M_{\rm M}$ that
is consistent with the more massive component of CM Dra.  This is
consistent, if slightly circular reasoning, but it suggests that the
inclination of the binary orbit plane is fairly high.  To avoid
eclipses, however, $i$ must not be too high, and thus higher stellar
masses are favored, in order to account for the observed radial
velocity amplitudes.

An alternate method of estimating the mass uses an empirical
mass-luminosity relation for M dwarfs, as follows.  From the white
dwarf's $T_{\rm eff}$ and $\log g$ and assuming $M_{\rm H} = 0$
($M_{\rm WD} = 0.40 M_{\odot}$) we find $\log L_{\rm WD}/L_{\sun}=-0.5$
(Althaus \& Benvenuto 1997\markcite{alt}), and thus $M_{\rm bol}({\rm
WD}) = 6.00$.  Using computed bolometric corrections and colors for
hydrogen atmosphere white dwarfs from Bergeron et al.\
(1995\markcite{berg}; for $T_{\rm eff}=30,000$~K and $\log g=8$), we
then find $M_V({\rm WD})=8.966$ and $M_R({\rm WD})=9.095$.  From the
spectral decomposition described above, the M star contributes
$\approx 16\%$ of the light in $R$, so $M_R({\rm M})= 10.90$.
According to Kirkpatrick \& McCarthy (1994\markcite{kirk}), an M4.5V
star has $V-R = 1.37$, so $M_V({\rm M}) = 12.27$.  Finally, Henry \&
McCarthy (1993)\markcite{hen} give the relation $\log M_{\rm
M}/M_{\sun}= -0.1681M_V+1.4217$ with an rms scatter in $\log
M/M_{\odot}$ of 0.081.  From this, $M_{\rm M}=0.23\pm
0.04\,M_{\odot}$.  The M star's mass inferred by this method is
slightly higher (by $\approx 0.02 M_{\sun}$) if the white dwarf's
hydrogen layer mass is $M_{\rm H} = 4 \times 10^{-4} M_{\sun}$,
because the higher white dwarf mass ($0.45 M_{\odot}$, Benvenuto \&
Althaus 1998\markcite{benalt}) implies a slightly larger radius for
the white dwarf (at fixed gravity), hence higher luminosities for both
stars. Either value of $M_{\rm M}$ is consistent with the orbit-based
mass estimate.

As an aside, we can compute the distance to the system.  The $R$
magnitude of the system at phase 0.0 (when the contribution of the
heated face of the M star is minimal) is $R=16.22$.  The white dwarf
contributes 84\% of the light in $R$, hence the apparent $R$ magnitude
of the white dwarf by itself is $m_R({\rm WD}) =16.22-2.5\log
0.84=16.41$.  The distance is then $d=290$ pc ($0.4 M_{\odot}$ white
dwarf) or $d=308$ pc ($0.45 M_{\odot}$ white dwarf).

\subsection{Light Curve Models}

The light curves described above show no obvious eclipses and are
nearly sinusoidal. They are thus consistent with a very simple picture
of the light modulation in PG 1224+309 being due to phase-dependent
visibility of the bright, inward-facing hemisphere of a nearly
spherical secondary star, which is uniformly emitting as a consequence
of strong heating by the white dwarf.  If one were sure of the
radiative properties of both illuminated and unilluminated sides of
the M dwarf (i.e., if one knew the surface fluxes and limb-darkening
coefficients in each color), one could model the light curves to infer
constraints on the radius of the M dwarf and on the inclination.  Such
radiative quantities are becoming available for normal, unilluminated
photospheres of low-luminosity stars (cf.\ 
Claret 1998\markcite{Claret} 
for
limb-darkening coefficients based on the Next-Gen model atmospheres of
Hauschildt et al.\ 1998\markcite{haus}).  
For the strongly illuminated hemisphere,
however, neither surface fluxes nor limb-darkening behavior is
known. Treating the problem by assuming that the atmosphere behaves
``normally,'' i.e.\ with standard albedo and limb-darkening
coefficients at an enhanced effective temperature, is not adequate for
making robust conclusions regarding the secondary star (cf.\
Hilditch et al.\ 1996\markcite{hild}, 
who review problems of light-curve synthesis for
strongly irradiated stars, including the formal finding of limb-{\it
brightening} in their analysis of the systems AA Dor and KV Vel).

In view of this difficulty, we are content here to check whether the
light curve amplitudes can be approximately reproduced, using
properties of the stars and the binary orbit consistent with our prior
discussion, and without requiring such an extreme inclination as to
introduce an eclipse or other large distortion.  Essentially, the
question reduces to whether the pair of quantities $i$ and $R_{\rm
M}/a$ that is needed to model the light curves is consistent with the
properties of PG 1224+309 as known from the discussion given above.
For this purpose, a simplified treatment using blackbody emissivities
(at the local effective temperature) and a limb-darkening law
appropriate for (unilluminated) cool photospheres {\it may} be
adequate.

We used the Wilson--Devinney (W--D; 1971\markcite{w-d}) 
light curve synthesis code,
fixing the temperature of the white dwarf at 29,300 K.  The M star's
temperature was fixed at 3100 K (unilluminated).  The gravity
darkening coefficients for the two stars were 1.0 and 0.3,
respectively. The bolometric albedos for reflection heating and
re-radiation were set to 1.0 and 0.5, respectively.  For definiteness,
a limb-darkening law was used that includes a logarithmic term and
took the appropriate coefficients for the four filters from the tables
in Van Hamme (1993\markcite{vanH}).  
In the W--D code, the fractional radii $R/a$ are
expressed in terms of the mass ratio and the so-called
$\Omega$-potentials of the two stars.  The inclination and mass ratio
were varied on a grid, $70^{\circ} \le i\le 79^{\circ}$ and $0.52\le
Q\le 0.71$.  For $i > 79^{\circ}$ eclipses are predicted.  At each
grid point, and assuming values for the $\Omega$-potentials, the $B$,
$V$, $R$, and $I$ light curves were fitted simultaneously, and the
values of $\Omega_{\rm WD}$ and $\Omega_{\rm M}$ were adjusted to find
the minimum $\chi^2$.  In general, the best-fitting models
underestimated the amplitude of the $B$ light curve, matched more or
less the amplitudes of the $V$ and $R$ light curves, and overestimated
the amplitude of the $I$ light curve.  In view of the modeling
uncertainties in the case of strongly illuminated stars, the ability
of the W--D code to roughly match the observed light-curve amplitudes
for plausible values of $i$, $Q$ and $\Omega_{\rm M}$ is regarded as
confirmation of the basic picture we have of PG 1224+309.

\subsection{Prospects for Improved Understanding of PG 1224+309}

Without the benefit of eclipses, how can accurate masses of the
component stars of PG 1224+309 be derived?  Additional observables
include the M star's rotational velocity $v_{\rm rot} \sin i$ and a direct
measurement of $K_{\rm M}$ from absorption lines.  The rotational
velocity could be exploited, assuming synchronous rotation, to form
the ratio $R_{\rm M}/a_{\rm M}$ where $a_{\rm M}$ is the size of the
M star's orbit around the center of mass.  An absorption line velocity
curve, depending on the lines used, would give an estimate of $K_{\rm M}$
requiring a ``K-correction'' of the opposite sign to that used here.

Additional information could be extracted from the light curves, given
improved theoretical modeling of irradiated atmospheres, to be
incorporated into light-curve synthesis codes.  In particular,
believable light-curve amplitudes could be used to specify the radius
of the M star, relative to the orbital separation, as a function of
the orbital inclination.  Consistency between inferred mass and radius
could then be demanded, to sharply limit the allowed range of inclination.

In principle, one could measure the gravitational redshift of the
white dwarf which would provide an independent constraint on the mass
and radius of the white dwarf.  The gravitational redshift would
simply be the difference in the two $\gamma$ values for the orbital
solutions of the two components.  Currently, the measured $\gamma$
values are not known well enough,  both in terms of formal statistical
errors and in terms of systematic errors, to enable a meaningful
measurement of the gravitational redshift ($\gamma_{\rm WD} = -4 \pm
12$ km s$^{-1}$ and $\gamma_{{\rm H}\alpha} = 10 \pm 6$ km s$^{-1}$;
statistical errors are quoted).  The Balmer lines have sharp emission
features in their cores, which complicates the measurement of
$\gamma_{\rm WD}$.  Possibly the white dwarf has metal absorption
lines in the ultraviolet, in which case a spaced-based measurement of
$\gamma_{\rm WD}$ might be possible.

Finally, some improvement in the measurement of $K_{\rm WD}$ might be
achieved from additional spectroscopy covering the entire orbital
cycle.

\subsection{Evolutionary Considerations}

\subsubsection{The Past: PG 1224+309 as a Post-Common Envelope Binary}

The present orbital separation of PG 1224+309 is $a \approx
1.5\,R_{\odot}$.  The present white dwarf was once the core of a giant
star with a radius of $\approx 100\,R_{\odot}$ and therefore $a >
100\,R_{\odot}$ formerly.  The standard picture of how such initially
wide binaries can be drastically reduced in orbital size is the
so-called ``common envelope'' (CE) process. Very roughly, in this
picture the envelope of the growing giant star engulfs the companion
star, and orbital energy is lost due to friction against the envelope,
leading to the possible ejection of the envelope.  
Iben \& Livio (1993\markcite{IL}) 
present a well-organized and relatively recent review of the CE
process and the stellar evolution scenarios and products involving it.
Livio (1996\markcite{liv}) 
reviews how observations can further constrain the CE
scenarios.

The CE phase is very short-lived, and it unlikely to be observed
directly, but post-CE binaries such as PG 1224+309 have the potential
to illuminate some aspects of the CE process. One question of interest
is whether a rule can be given for terminating the decay of the binary
orbit --- does the decay continue until essentially all material
outside the (pre-white dwarf) ``core'' of the giant star has been
expelled, and if so what is the mass of the residual hydrogen
envelope?  In PG 1224+309, $M_{\rm WD}$ as inferred from $T_{\rm eff}$
and $\log g$ is ambiguous at the $0.05~M_{\sun}$ level, unless the
hydrogen envelope mass $M_{\rm H}$ is known.  Given an estimate of
$M_{\rm WD}$, however, we can infer $M_{\rm M}$ along with the orbital
inclination, from $\Delta K$ and the mass ratio.  With our best
estimate of $\Delta K$, the lack of eclipses favors a higher white
dwarf mass, hence a relatively thick hydrogen envelope. PG 1224+309
misses being an eclipsing system by only a few degrees! Eclipses would
provide additional constraints on the mass and radius of the white
dwarf.  It would then be possible to infer $M_{\rm H}$ with high
confidence.  This would be a new and valuable piece of information
with which to confront common-envelope theories.

Systems like PG 1224+309 can also sometimes be used to study the
efficiency of the common envelope process.  This is usually
parameterized by the parameter $\alpha_{\rm CE}$, which relates the
initial and final orbits to the energy expended in ejecting the CE.
The efficiency thus helps to determine the mapping of initial binary
periods onto final (post-CE) periods.  Except for the ambiguity in
$M_{\rm WD}$, PG1224+309 would be a useful system in potentially
giving a lower limit to $\alpha_{\rm CE}$.  The argument is as
follows. The decay of the orbit must have released enough energy to
overcome the binding energy of the envelope that formerly blanketed
the white dwarf.  The supply of orbital energy can be estimated from
the present separation and masses (there is little dependence on the
initial separation), and the present mass of the white dwarf gives the
radius of its giant star progenitor at the moment that dynamical
timescale mass transfer began ({\it via\/} the giant branch core-mass
-- luminosity relation).  That radius, combined with an estimate of
the envelope mass, gives the binding energy of the envelope.  The
ratio defines $\alpha_{\rm CE}$: $ E_{\rm bind} = \alpha_{\rm CE} \Delta
E_{\rm orbit} $ or
$${M_1(M_1-M_{1R}) \over {\lambda a_o r_L}} = 
\alpha_{\rm CE} \left(
{M_{1R} M_2 \over {2a_f}} - {M_1 M_2 \over {2a_o}}
\right)$$
using the definitions of 
Webbink (1984\markcite{webb}) as given in 
Livio (1996\markcite{liv}).
Here $M_1$ and $M_2$ are the initial masses, $M_{1R}$ is the core or
remnant mass of the more evolved star, $a_o$ and $a_f$ are the
original and final orbital separations, $r_L$ is the radius of the
evolved star's original Roche lobe expressed as a fraction of $a_o$,
and $\lambda = 0.5$ is a structure factor for that star's envelope.

The calculated value of $\alpha_{\rm CE}$ depends strongly on the
assumed core mass $M_{1R}=M_{\rm WD}$.  For PG 1224+309, we first
consider the case where $M_{1R} = M_{\rm WD} = 0.40\,M_{\sun}$, $M_2 =
0.25\,M_{\sun}$, and $a_f = 1.48\,R_{\sun}$. Then the available
orbital energy was about $\Delta E_{\rm orbit} =3.4 \times 10^{-2}$ in
solar units with $G=1$.  Here we neglect the small second term on the
R.H.S. of the displayed equation. For assumed initial masses of $M_1 =
1.0$, 1.2, and 1.4 $M_{\sun}$, the radius of the star when it
contacted its Roche lobe and began mass transfer leading to the common
envelope was $R_{\rm lobe} = a_o r_L = 75$, 71, or 68~$R_{\sun}$
(using approximations from Tout et al.\ 1997\markcite{tout}). The
binding energy of the envelope was approximately $E_{\rm bind} = 1.6
\times 10^{-2}$, $2.7 \times 10^{-2}$, and $4.1 \times 10^{-2}$,
respectively.  With these assumptions, the available orbital energy
exceeded the binding energy only if $M_1$ was less than
1.4~$M_{\sun}$, and even then most of the orbital energy was needed to
unbind the envelope; thus $\alpha_{\rm CE}$ cannot have been
significantly less then unity.  (Stars with values of $M_1$ much
smaller than those considered are less likely to have evolved so far
yet. Also, for $M_{1R} = 0.4\,M_{\sun}$, which corresponds to a
luminosity well below the tip of the red giant branch, mass loss on
the giant branch is not likely to have been significant.)  By this
example we have shown how, in principle, observations of relatively
wide post-CE binaries (with small negative orbital energies) set a
lower limit on $\alpha_{\rm CE}$.  However, in the specific case of PG
1224+309, the ambiguity in $M_{\rm WD}$ owing to our ignorance of
$M_{\rm H}$ makes this particular test less interesting. This is
because increasing $M_{\rm WD}$ to $\approx 0.45\,M_{\sun}$ reduces
the pre-CE binding energy greatly and thus reduces the lower limit on
$\alpha_{\rm CE}$.  Finding {\it eclipsing\/}, double-lined DA+dM
systems may in favorable cases provide firmer constraints on
$\alpha_{\rm CE}$.  Analogous eclipsing systems would allow better
constraints to be placed on the component masses {\em and} radii,
which would in turn allow one to begin addressing additional
questions, such as whether post-CE secondaries satisfy a normal main
sequence mass-radius relation.

\subsubsection{The Future: PG 1224+309 as a Pre-Cataclysmic Binary}

PG 1224+309 is a post-CE binary, as discussed above, and is also a
pre-cataclysmic binary in the sense of 
Ritter (1986\markcite{ritter}): the orbital
separation $a$ will shrink owing to gravitation wave radiation (GWR)
until the M star contacts its Roche lobe and initiates mass transfer
onto the white dwarf.  (Angular momentum loss {\it via\/} magnetic
braking, another process considered by Ritter, is thought not to
operate for low-mass M dwarfs.) Using formulae from 
Ritter (1986\markcite{ritter}) and
adopting $M_{\rm WD} = 0.45\,M_{\sun}$ and $M_{\rm M} =
0.28\,M_{\sun}$, it will take about $9.0 \times 10^9$~years for GWR to
bring PG 1224+309 into a semidetached configuration, at which time the
orbital period will be 2.5 hours.  The present cooling age of the
white dwarf is perhaps a few $\times 10^7$~years 
(Althaus \& Benvenuto 1997\markcite{alt}), 
but when Roche-lobe overflow (RLOF) mass transfer begins the
white dwarf will have cooled to well below $T_{\rm eff} = 5000$~K.

\section{Summary}

We have measured the orbital period of PG 1224+309 to be
$P=0.258689\pm 0.000004$ days.  The semi-amplitude of the white
dwarf's radial velocity curve is $K_{\rm WD}=112\pm 14$ km s$^{-1}$.
The velocity semi-amplitude of the H$\alpha$ emission line arising
from the irradiated hemisphere of the M star is $K_{{\rm
H}\alpha}=160\pm 8$ km s$^{-1}$.  We estimate a ``K-correction'' of
$\Delta K=21\pm 2$ km s$^{-1}$, where $K_{\rm M}=K_{{\rm
H}\alpha}+\Delta K$.  The implied mass ratio is then $Q=M_{\rm
M}/M_{\rm WD}=0.62\pm 0.08$.  The spectral type of the secondary is
M4+.  We fit synthetic spectra to the blue spectrum and find $T_{\rm
eff}=29,300$ K and $\log g=7.38$ for the white dwarf.  The white dwarf
contributes 97\% of the light at 4500~\AA\ and virtually all of the
light blueward of 3800~\AA.

The mass inferred for the white dwarf depends on the assumed mass of
the thin residual hydrogen envelope: $0.40\le M_{\rm WD}\le
0.45\,M_{\sun}$ for hydrogen envelope masses in the range $0\le M_{\rm
H} \le 4\times 10^{-4}\,M_{\odot}$.  Given the absence of eclipses,
for likely values of $\Delta K$ and based on the observed values of
$K_{\rm WD}$ and $K_{\rm M}$, we conclude that the mass of the white
dwarf is closer to $0.45\,M_{\odot}$. It thus appears that the white
dwarf has a relatively large residual hydrogen envelope.  The mass of
the M star is $M_{\rm M}=0.28\pm 0.05\,M_{\odot}$ if $\Delta K=21\pm
2$ km s$^{-1}$ and $M_{\rm WD}=0.45\pm 0.05\,M_{\odot}$.  In this
case, the inclination is $i=77\pm 7^{\circ}$, only a few degrees away
from giving rise to eclipses.  Additional observational and
theoretical work could improve our knowledge of the masses of the
component stars.

We argued that an accurately determined residual hydrogen envelope
mass can constrain theories of common envelope evolution, and we showed
how accurate values of $M_{\rm WD}$ in binaries such as PG 1224+309
can help set limits on $\alpha_{\rm CE}$, the efficiency parameter used in
common envelope scenarios.  PG 1224+309 itself will become a
cataclysmic variable with an orbital period of about 2.5 hours, after
a further $\approx 10^{10}$~years of orbital evolution.

\acknowledgements

This work was partially supported by a grant from NASA administered by
the American Astronomical Society.  JRT and CJT thank the NSF for
support through grant AST-9314787, and the MDM staff for their usual
excellent support.  M.\ E.\ acknowledges support from Hubble fellowship
grant HF-01068.01-94A from Space Telescope Science Institute, which is
operated for NASA by the Association of Universities for Research in
Astronomy, Inc., under contract NAS~5-26255.  We thank Ivan Hubeny for
instruction in the use of his stellar atmosphere codes {\sc TLUSTY},
{\sc SYNSPEC}, and {\sc ROTINS}.

\end{document}